\documentclass[%
superscriptaddress,
 amsmath,amssymb,
 aps,
 prl,
 twocolumn,
]{revtex4-2}
\usepackage{CJK}
\usepackage{graphicx}
\usepackage{dcolumn}
\usepackage{bm}
\usepackage{graphicx}
\usepackage{amsmath}
\usepackage{xcolor}
\usepackage{booktabs}

\begin{document}
\begin{CJK*}{UTF8}{gbsn}

\preprint{APS/123-QED}

\title{Fine particle percolation in a sheared granular bed}

\author{Song Gao (高颂)}
\affiliation{Department of Mechanical Engineering}

\author{Julio M. Ottino}
\affiliation{Department of Mechanical Engineering}
\affiliation{Department of Chemical and Biological Engineering}
\affiliation{Northwestern Institute on Complex Systems (NICO), Northwestern University, Evanston, Illinois 60208, USA}

\author{Richard M. Lueptow}
\affiliation{Department of Mechanical Engineering}
\affiliation{Department of Chemical and Biological Engineering}
\affiliation{Northwestern Institute on Complex Systems (NICO), Northwestern University, Evanston, Illinois 60208, USA}

\author{Paul B. Umbanhowar}
\email{umbanhowar@northwestern.edu}
\affiliation{Department of Mechanical Engineering}

\date{\today}

\begin{abstract}
We study the percolation velocity, $v_p,$ of a fine spherical particle in a sheared large-particle bed under gravity using discrete element method simulations for large-to-fine particle diameter ratios, $R=d/d_f,$ below and above the free-sifting threshold, $R_t\approx6.5.$ For $R<R_t,$ $v_p$ initially increases with increasing shear rate, $\dot\gamma,$ as shear-driven bed rearrangement reduces fine-particle trapping but then decreases toward zero due to fine-particle excitation for $\dot\gamma\sqrt{d/g}\gtrsim 0.1$. For $R>R_t$, $v_p$ is constant at low $\dot\gamma$ but decreases toward zero at higher shear rates due to fine-particle excitation.
\end{abstract}

\maketitle
\end{CJK*}

Mixtures of granular materials often segregate due to differences in size, density, or other physical properties, and understanding this phenomenon is often critical for predicting and controlling various natural~\citep{iverson1997physics,frey2009river,johnson2012grain} and industrial processes~\citep{ottino2000mixing,ottino2008mixing}. Recent studies have advanced the modeling of granular segregation~\citep{gray2018particle,umbanhowar2019modeling}, with size segregation in dense granular flows receiving the majority of attention~\citep{bridgwater1976fundamental,savage1988particle,gray2005theory,johnson2012grain,marks2012grainsize,fan2014modelling,tunuguntla2014mixture,schlick2015modeling,trewhela2021experimental}. Nearly all of these studies consider mixtures with large-to-small particle diameter ratios, $R=d/d_f \lesssim 2,$ where interparticle contacts are enduring~\citep{marks2012grainsize,fan2014modelling,tunuguntla2014mixture}. In these cases, segregation can be characterized by a concentration-dependent \emph{percolation velocity}, $v_p$, which is typically predicted and observed to increase monotonically with both $R$ and the shear rate, $\dot{\gamma}.$ 

For larger $R$, where small particles are referred to as \emph{``fine,"} the $v_p$-dependence on $\dot\gamma$ and $R$ in sheared flows changes significantly. In particular, for $R\gtrsim 2$ and low fine-particle concentration, $v_p$ increases dramatically with increasing $R$~\citep{scott1975interparticle,trewhela2021experimental}, but $v_p$ is nearly $R$-independent at larger fine-particle concentration (above 10\%) for $2\lesssim R \lesssim 4$~\citep{schlick2015modeling}. Here we focus on fine-particle segregation in uniform shear flow  in the zero-concentration limit where an increasing tendency toward free sifting (or spontaneous percolation), with increasing $R$, leads to qualitative changes in the dependence of $v_p$ on $\dot\gamma$ and other parameters. 

Free sifting has been investigated primarily in static beds for $R>R_t$~\citep{bridgwater1969particle,ghidaglia1996transition,ippolito2000diffusion,lomine2009dispersion,remond2010simulation}, where the free-sifting threshold, $R_t,$ is $R_{t0}=(2/\sqrt{3}-1)^{-1}\approx 6.46$ for rigid monodisperse spheres~\citep{dodds1980porosity} but is larger in polydisperse mixtures of ``soft"  particles.  Free sifting can also occur for $R<R_t$ in randomly packed static beds when a sub-population of pore throats---the minimum opening between neighboring bed spheres---exceeds the fine-particle diameter. In this case, percolation is necessarily \emph{transient} since a fine particle will inevitably encounter an impassible pore throat~\citep{gao2023percolation}.  Despite the ubiquity of fine particles in industrial solids processing~\citep{bemrose1987review,schulze2008powders}, their potential for increasing the mobility of various geophysical flows~\citep{phillips2006enhanced,linares2007increased,chassagne2020mobility}, and their importance in sediment infiltration that shapes river dynamics, channel morphology and ecological habitats~\citep{dietrich1989sediment,wilcock1998two}, few studies have focused on fine-particle percolation velocity in granular shear flows~\citep{scott1975interparticle,khola2016correlations,trewhela2021experimental}.

We show here that free sifting is pronounced and unavoidably coupled with shear for $R<R_t,$ because fines that would be trapped in a static bed are repeatedly re-mobilized by shear-induced particle rearrangements. In past work, the complexity of this problem and the limited parameter-space explored, produced puzzling inconsistencies regarding the dependence of $v_p$ on $\dot{\gamma}$~\citep{khola2016correlations} and $R$~\citep{scott1975interparticle,trewhela2021experimental,schlick2015modeling}. In this Letter we resolve these issues by characterizing the fine-particle percolation velocity in large-particle beds for size ratios spanning the free-sifting threshold ($2\leq R \leq 10$) and spatially-uniform shear rates covering the  quasi-static and rapid dense flow regimes~\citep{tardos2003slow}. Our results reveal a non-monotonic dependence of $v_p$ on $\dot\gamma$, provide relations for predicting $v_p$ in the low- and high-shear rate regimes, and add insight into the dominant physics in each regime.

\emph{Methods}---LIGGGHTS~\cite{kloss2012models}, a discrete element method code, is used to simulate single fine-particle percolation in a confined dense granular flow with a prescribed linear velocity profile~\cite{fry2018effect}. The flow domain is periodic in the streamwise and spanwise directions and confined in the depthwise ($y$) direction by two horizontal planar walls roughened by randomly attached bed particles.  A constant downward force on the top wall, which is otherwise free to move vertically and in the spanwise direction, sets the bed overburden pressure, $P_0$, which is increased with increasing $\dot\gamma$ to maintain a constant volume fraction $\phi\approx 0.58$ of bed particles. The bottom wall is stationary while the top wall is translated in the streamwise direction with velocity $\dot{\gamma}h(t),$ where the time-dependent bed height, $h(t),$ accommodates dilation due to shear.  Parameters are set as follows: bed-particle diameter $d=5$\,mm with 10\% uniform polydispersity, gravitational acceleration $g=9.81$\,m\,s$^{-2}$ (in the $-y$ direction), restitution coefficient $e=0.8$, friction coefficient $\mu=0.5$, and bed- and fine-particle densities of $\rho=\rho_f=2500$\,kg\,m$^{-3}$. We also vary $d$, $g$, $e$, and $\rho_f$ to explore their effects on $v_p$. Depending on $R$, between $\sim10^3$ ($R=2$) and $\sim10^4$ ($R=10$) single fine particles with initial streamwise velocity matching the porous upper moving wall are dropped into the sheared bed. Fine particles interact with bed particles but not with each other to examine the zero-concentration limit.

\begin{figure}[t]
\centering
\includegraphics[scale=0.48]{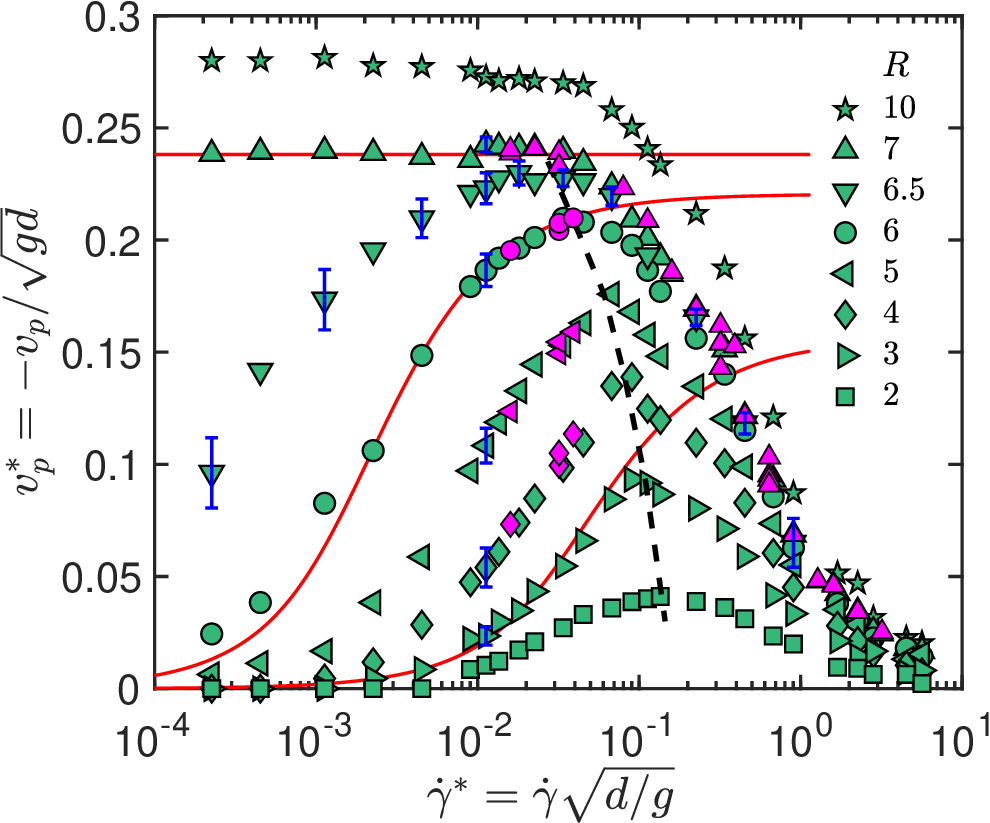}
\caption
{\label{fig:fig1} Scaled fine-particle percolation velocity, $v_p^*=-v_p/\sqrt{gd},$ vs.\ scaled shear rate, $\dot\gamma^*=\dot\gamma\sqrt{d/g},$ for various particle size ratios, $R$, gravitational accelerations, $g$, and bed particle diameters, $d$, with restitution coefficient $e=0.8$. Volume fraction of large particles is kept nearly constant ($0.57<\phi<0.58$) by increasing the overburden pressure, $P_0,$ as $\dot\gamma$ is increased. Magenta symbols indicate data for $d\in\{2.5, 10, 15\}$\,mm and $g\in\{4.91, 19.62, 29.43\}$\,m/s$^2$ at $\dot{\gamma}=1$\,s$^{-1}$ for $R\leq6$ and for $1 \leq \dot\gamma \leq 100$\,s$^{-1}$ for $R=7$. Dashed black curve approximates maximum $v_p^*$ location for different $R$. Error bars for selected cases indicate standard error. Red curves are predictions of the low-shear-rate regime model (Eq.~\ref{eq2}) for $R=3$, 6, and 7.}
\end{figure}

\emph{Percolation velocity}---Figure~\ref{fig:fig1} plots the scaled fine-particle percolation velocity, $v_p^*=-v_p/\sqrt{gd},$ versus scaled shear rate, $\dot\gamma^*=\dot\gamma\sqrt{d/g},$ for $2 \leq R \leq 10$.  As in static beds, single fine particles always percolate downward on average even at the largest $\dot\gamma^*$, and $v_p^*$ increases monotonically with increasing $R$ for all $\dot\gamma^*$. However, the dependence of $v_p^*$ on $\dot\gamma^*$ is strongly $R$-dependent.  First, for size ratios corresponding to the static-bed passing regime ($R>6.5>R_{t0}$ here due to overburden-pressure-driven deformation and polydispersity of bed particles that decreases the minimum pore throat diameter relative to rigid monodisperse bed particles), $v_p^*$ initially remains constant at its static-bed value as $\dot\gamma^*$ is increased from zero. Hence, $v_p\propto \sqrt{gd}$ for $R\gtrsim R_t$. However, for $\dot\gamma^* \gtrsim 0.03$, $v_p^*$ decreases with increasing $\dot\gamma^*$. Second, for size ratios in the static-bed trapping regime ($R\leq 6.5$), $v_p^*$ increases from zero with increasing $\dot\gamma^*$, similar to segregation with $R\lesssim 2$~\citep{marks2012grainsize,fan2014modelling,tunuguntla2014mixture,trewhela2021experimental}. However, $v_p^*$ reaches a maximum near $\dot\gamma^*\sim 0.1$ and then decreases toward zero with further increase in $\dot\gamma^*.$  Note that the previously observed $\dot{\gamma}$-independence of $v_p$ for $R\approx2.5$~\citep{khola2016correlations} results from that study's limited shear rate range, $0.04<\dot{\gamma}^*<0.14$, which brackets the peak in $v_p$ and about which $v_p$ is nearly constant (e.g., see $R=2$ in Fig.~\ref{fig:fig1}). To test the nondimensionalization of $v_p$ and $\dot\gamma$, Fig.~\ref{fig:fig1} also includes data where $d$ and $g$ differ from the values used in the other simulations. This additional data (magenta symbols) overlays the data for $d=5$\,mm and $g=9.81$\,m/s$^2$ at the corresponding $R$ values, indicating that the scaling is correct in both low and high $\dot\gamma$ regimes. 

The value of $\dot\gamma^*$ where $v_p^*$ begins to drop decreases with increasing $R$ (dashed curve), e.g., $\dot{\gamma^*}\approx0.14$ for $R=2$, while  $\dot{\gamma^*}\approx0.03$ for $R=6.5$. This sensitivity to $R$ along with the decrease in $v_p$ with increasing $\dot\gamma$ for $\dot\gamma^* \gtrsim 0.1$ is due to increasing fine-particle velocity fluctuations (here characterized by fluctuations in the vertical velocity component, $v_\mathrm{rms}$), which frustrate percolation and increase with increasing $R$ or $\dot\gamma^*$, as described later. In static beds, a similar decrease in $v_p$ is observed with increasing $e$ due to velocity fluctuations for both $4\leq R\leq R_t$~\citep{gao2023percolation} and for $R>R_t$~\cite{bridgwater1969particle,lomine2009dispersion,ippolito2000diffusion, gao2023percolation}, as fine particles rebound more energetically after colliding with bed particles.

Figure~\ref{fig:fig1} and previous work in static beds~\cite{gao2023percolation} suggest that fine-particle percolation in sheared beds depends on three mechanisms: geometric trapping, which is possible when $R<R_t$; bed particle rearrangement due to shear; and fine-particle velocity fluctuations, which frustrate percolation. Details of their contributions to fine-particle transport are developed below, but the basics are as follows: First, when fine particles with $R<R_t$ are trapped, they re-mobilize due to shear-driven bed rearrangements at a rate that is proportional to $\dot\gamma$ and increases with $R$, indicating that passable voids are generated at a higher rate for smaller fine particles. Second, the average time to pass through a void increases with increasing excitation of the fine particles, measured in terms of $v_\mathrm{rms}$. Consequently, at high shear rates, where trapping times are short for $R<R_t$ and $v_\mathrm{rms}$ is large ($v_\mathrm{rms}\propto \dot\gamma$) for all $R$, $v_p$ decreases with increasing $\dot\gamma$.

\emph{Low-shear-rate regime}---To better understand the dominant physics and develop a model for $v_p$ in this regime, we start with the percolation depth model for $R<R_t$ in static beds, $p(\Delta y)\propto P_p^{\frac{\Delta y}{d}}$, where $p(\Delta y)$ is the probability that a fine particle falls a distance $\Delta y$ or more from its starting height, and $P_p$ represents the probability of a fine particle passing through a randomly selected pore throat, i.e., the fraction of constrictions with diameters larger than $d_f$~\citep{gao2023percolation}. In static beds, $p(\Delta y)$ is the proportion of trapped fine particles that exceed a depth $\geq \Delta y$, assuming that the passage of fine particles through consecutive pore throats is independent~\citep{meakin1990simulation,gao2023percolation}. Since untrapped fine particles percolate with mean velocity $v_{p,s}=-c_1\sqrt{gd}$~\citep{gao2023percolation}, $p(\Delta y)$ can be reformulated as a function of time using $\Delta y=-v_{p,s}t$ as $p(t)\propto P_p^{\frac{-v_{p,s}t}{d}},$ where $p(t)$ is the probability that a fine particle is untrapped after time $t.$ The average percolation velocity over $t$ is then 
\begin{equation} \label{eq1}
v_p=\frac{v_{p,s}}{t}\int_{0}^{t} p(t')\,dt'\propto \frac{d}{t\ln{P_p}}\left(1-P_p^{\frac{-v_{p,s}t}{d}}\right).
\end{equation}

For sheared systems, we assume that the time interval between significant bed rearrangements scales as $t_b=c_2\dot\gamma^{-1}$, where $c_2$ depends on $R$. Substituting $t_b$ for $t$ in Eq.~\ref{eq1} gives $v_p$ as a function of shear rate, bed structure (via $P_p$) and its variation (via $c_2$), bed particle diameter, and gravitational acceleration:
\begin{equation} \label{eq2}
v_p= \frac{d\dot\gamma}{c_2\ln{P_p}}\left(1-P_p^{c_1c_2\sqrt{\frac{g}{\dot\gamma^2d}}}\right).
\end{equation}
This relation is alternatively expressed as
\begin{equation} \label{eq3}
v_p^{\prime}= \frac{v_p}{v_{p,s}}=\dot{\gamma}^\prime\left(1-e^{-1/\dot{\gamma}^\prime}\right),
\end{equation}
where $\dot{\gamma}^\prime=-C\dot{\gamma}\sqrt{d/g}$ with $C^{-1}=c_1c_2\ln{P_p}$ as the single model parameter. Eq.~\ref{eq2} exhibits the appropriate limiting behaviors under its assumption that velocity fluctuations are small: i) as $\dot{\gamma}\to \infty$ ($t_b \to 0$), $v_p\to v_{p,s}\propto \sqrt{gd}$ for all $R$; ii) as $\dot{\gamma}\to 0$, $v_p\propto d\dot{\gamma}$ for the trapping regime ($R<R_t$) as in most shear-driven percolation models for small $R$~\citep{marks2012grainsize,fan2014modelling,tunuguntla2014mixture,trewhela2021experimental} and is independent of $g$; iii) in the passing regime ($R>R_t$, $P_p=1$) $v_p \propto \sqrt{gd}$ independent of $\dot\gamma$ as in i).

\begin{figure}[t]
\centering
\includegraphics[scale=0.8]{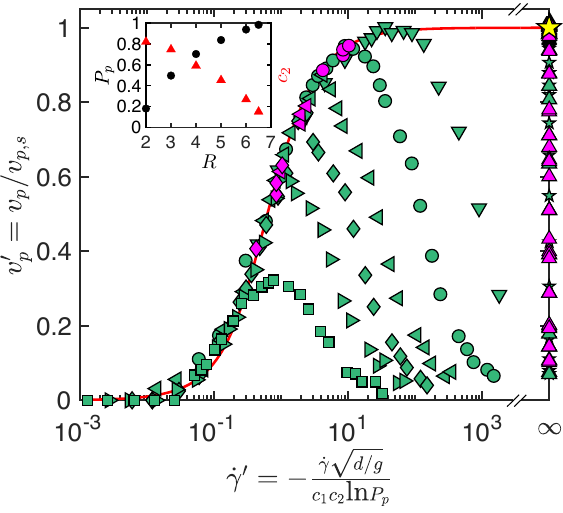}
\caption
{\label{fig:fig2} Fine-particle percolation velocity scaled by percolation velocity of untrapped fine particles in static bed,  $v_p/v_{p,s},$ vs.\ rescaled shear rate, $\dot{\gamma}^\prime,$ showing collapse of all data (symbols) from Fig.~\ref{fig:fig1} in the low-shear-rate-regime onto the prediction  of Eq.~\ref{eq3} (red curve).  Data include varying $R$ (symbols), and $g$ and $d$ (magenta) for $\phi\approx0.58$ and $e=0.8$. Passing regime data ($R>6.5$) fall on the right boundary since $\dot{\gamma}^\prime=\infty$. Inset: $P_p$ (left: black circle) and $c_2$ (right: red triangle) vs.\ $R$.}
\end{figure}

To compare Eq.~\ref{eq2} to our data, we determine $P_p$ by characterizing the pore throat size distribution using Delaunay triangulation~\citep{al2003comparison,reboul2008statistical,gao2023percolation}. For $\phi\approx 0.58$, $P_p$ is nearly independent of shear rate for $\dot{\gamma}\lesssim0.1$, and increases from 0.17 to 0.93 as $R$ is increased from 2 to 6. From~\citep{gao2023percolation}, $c_1=0.09\sqrt{R}$ for $\phi\approx0.58$ and $e=0.8$. Fits of Eq.~\ref{eq2} to simulation results for three $R$ values obtained by adjusting the one free parameter, $c_2,$ match the simulation data at low $\dot\gamma^*$, as shown in Fig.~\ref{fig:fig1} (solid curves). The inset in Fig.~\ref{fig:fig2} shows that $P_p$ increases with $R$ and $c_2$ decreases with $R$, as would be expected.
 
All data in Fig.~\ref{fig:fig1} is compared to the universal form of the model (Eq.~\ref{eq3}) in Fig.~\ref{fig:fig2}, which plots the percolation velocity scaled by the untrapped percolation velocity from the static bed, $v_p' = v_p/v_{p,s},$ versus the rescaled shear rate $\dot\gamma' = -C \dot\gamma^*$. Data for all $R$ as well as varying $g$ and $d$ (magenta) collapse onto the model prediction (red curve) in the low-shear-rate region. Note that for free-sifting cases ($R>6.5$), $\dot\gamma'=\infty$ since $P_p=1$, and the corresponding symbols fall on the far right of Fig.~\ref{fig:fig2} and go to $v_p'=1$ (yellow star) in the low-shear-rate regime.

\begin{figure*}[t]
\includegraphics[scale=0.74]{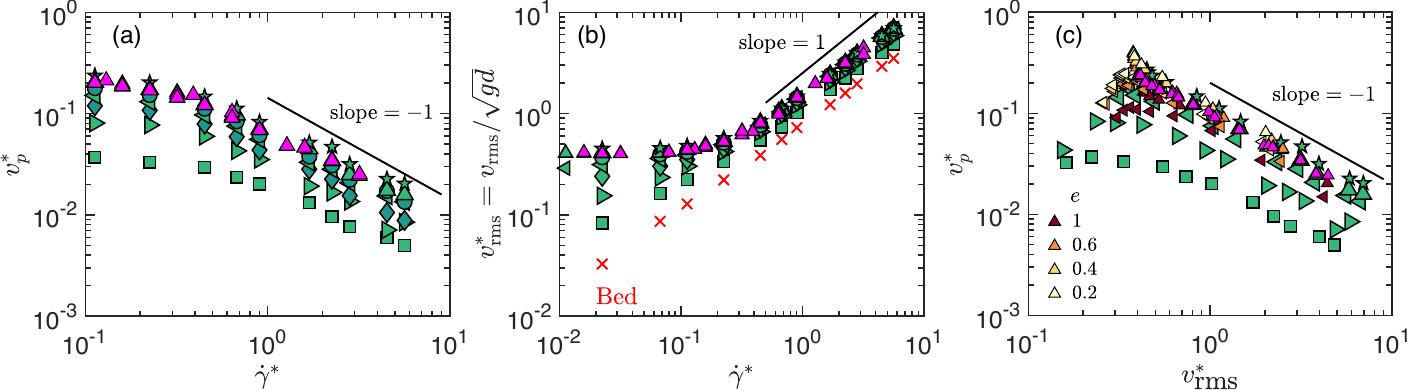}
\caption{
\label{fig:fig3} 
(a) Scaled fine-particle percolation velocity, $v_p^*=-v_p/\sqrt{g d}$, varies inversely with scaled shear rate, $\dot\gamma^*=\dot\gamma\sqrt{d/g}$. (b) Scaled fine-particle RMS velocity fluctuations, $v_\mathrm{rms}^*=v_\mathrm{rms}/\sqrt{gd}$, increase linearly with $\dot\gamma^*$ for $\dot{\gamma}^* \gtrsim 0.4$. Bed particle data ($\color{red}\times$) shown for comparison. (c) $v_p^*$ varies inversely with $v_\mathrm{rms}^*$ for various shear rates, and restitution coefficients, $e$ (colors as indicated). Data in (a-c) includes various size ratios $2\leq R\leq10$ (symbols), and $g$ and $d$ values (magenta) as in Fig.~\ref{fig:fig1}. 
}
\end{figure*}

\emph{High-shear-rate regime}---When $v_p^*$ for $\dot\gamma^* \gtrsim 0.1$ is plotted versus $\dot\gamma^*$ on a log-scale in Fig.~\ref{fig:fig3}(a), it is clear that $v_p^* \propto 1/\dot\gamma^*$ for $\dot\gamma^*\gtrsim 0.4$ and different $R$, $g$ and $d$. This behavior is related to increasing fine-particle velocity fluctuations, which frustrate percolation. To demonstrate the relation between $\dot\gamma^*$ and $v_\mathrm{rms}$, we first plot the scaled vertical root-mean-square velocity fluctuations of fine particles, $v_\mathrm{rms}^*=v_\mathrm{rms}/\sqrt{gd}$, versus $\dot{\gamma}^*$ for various $R$ in Fig.~\ref{fig:fig3}(b). For context, the  vertical velocity fluctuations of \emph{bed} particles ($\color{red} \times$) increase linearly with $\dot{\gamma}^*$, as would be expected from the corresponding increase in inter-particle collisions. Similarly, for $\dot{\gamma}^*\gtrsim 0.4$, $v_\mathrm{rms}^*\propto \dot\gamma^*$ for all $R$, indicating that fine-particle velocity fluctuations are linked to the bed-particle velocity fluctuations in the high $\dot\gamma^*$ regime. In comparison, for $\dot{\gamma}^*\lesssim 0.1$, gravity-driven fluctuations dominate, so that $v_\mathrm{rms}^*$ is either constant (free-sifting regime, $R>6.5$) or decreases slower than $\dot\gamma^*$ (trapping regime, $R\leq 6.5$). Simulations with different $g$ and $d$ values at $R=7$ [magenta triangles in Fig.~\ref{fig:fig3}(b)] confirm the scaling between $v_\mathrm{rms}$ and $\dot{\gamma}$, which indicates that $v_\mathrm{rms}$ is gravity independent where $v_\mathrm{rms}^* \propto \dot\gamma^*$ but proportional to $\sqrt{gd}$ where $v_\mathrm{rms}^*$ is constant.  

For all $\dot\gamma^*$, $v_\mathrm{rms}^*$ is always larger for larger $R$ and appears to approach a limiting curve for large $R$, as Fig.~\ref{fig:fig3}(b) shows. Additional simulations with fine-particle density varied by two orders of magnitude ($250$\,kg\,m$^{-3}$ to $2.5\times10^4$\,kg\,m$^{-3}$) at constant $R$ change $v_\mathrm{rms}$ by $<7\%$ ($v_p$ is also minimally affected), thereby indicating that the increase in $v_\mathrm{rms}$ with increased $R$ is due to decreased fine-particle diameter (i.e., smaller fine particles are less constrained by bed particles than larger fine particles) rather than decreased fine-particle mass. 

Having demonstrated the linear dependence of $v_\mathrm{rms}$ on shear rate at high $\dot\gamma$, Fig.~\ref{fig:fig3}(c) tests our hypothesis that the percolation velocity decreases with increasing $v_\mathrm{rms}$.  Indeed, the data show that $v_p^* \propto 1/v_\mathrm{rms}^*$ when
$v_\mathrm{rms}^*\gtrsim0.4$ for all $R$. In dimensional form, $v_p \propto gd/v_\mathrm{rms} \propto g/\dot\gamma,$ where the linear dependence of $v_p$ versus $\dot\gamma$ on $g$ alone at high $\dot\gamma$ contrasts with the low-shear-rate scaling of $v_p$ with $\sqrt{gd}$ in the free-sifting regime and with $d$ alone in the trapping regime.
Equally significant, the figure also includes data for varying restitution coefficient between bed and fine particles, $0.2\leq e\leq1$ for $R=5$ and 7, indicating that different combinations of $e$ and $\dot\gamma$ producing the same $v_\mathrm{rms}$ yield the same $v_p$.  Hence, it is $v_\mathrm{rms}$ that determines $v_p$ in this regime.

The $v_p^* \propto 1/v_\mathrm{rms}^*$ relationship can be understood in terms of the different rates at which fine particles exit bed particle voids traveling down versus up. This process mimics gas molecules escaping through a hole smaller than their mean free path, which is described by Graham's law of effusion for the flux, $\Phi = P_gA/\sqrt{2\pi m k_B T},$ where $m$ is the molecular mass, $P_g$ is the gas pressure, $A$ is hole area, and $T$ is the temperature. The analogy follows by replacing $P_g A$ with the gravity induced force differential in a trapping void, $mg,$ and $k_B T$ with $m v_\mathrm{rms}^2$. Multiplying $\Phi$ by the characteristic length $d$ to form a velocity gives $v_p \propto gd/v_\mathrm{rms}$, which is the dimensional form of the scaling shown in Fig.~\ref{fig:fig3}(c). A second analogy is the Drude model for electron transport in metals due to an electric field, $E$, in which the mean electron momentum is $p=m_e v=qE\tau,$ where $m_e$ and $q$ are the electron mass and charge, and $\tau$ is the characteristic time interval between collisions with heavier ions. Replacing $v$ with $v_p$, $q E/m_e$ with $g$, and $\tau$ with $d/v_\mathrm{rms}$ (since the fine-particle mean free path is proportional to $d$) yields $v_p \propto g d/v_\mathrm{rms}$.

\emph{Discussion}---Our simulations of gravity-driven percolation of single fine particles in sheared granular beds display different dominant physics at low and high shear rates. For low shear rates, $\dot\gamma\sqrt{d/g}\lesssim 0.1$, as the shear rate increases from zero, bed-particle rearrangements due to shear reduce fine-particle trapping to increase the percolation velocity, $v_p$. A statistical model of this mechanism (Eqs.~\ref{eq2} or~\ref{eq3}) accurately predicts $v_p$ for a wide range of conditions. In the high-shear-rate regime, $\dot\gamma\sqrt{d/g}\gtrsim 0.1$, increasing $\dot\gamma$ results in increasing fine-particle velocity fluctuations which frustrate percolation such that $v_p$ is inversely proportional to the velocity fluctuations and thus inversely proportional to $\dot\gamma$, as is evident from Fig.~\ref{fig:fig3}.

These results are far from complete, and suggest that many interesting questions and challenges remain. For instance, our model and scalings accurately capture the dependence of $v_p$ on $\dot\gamma$, $d$, and $g$, but require additional inputs to describe the effects of restitution coefficient, size ratio, and volume fraction. Understanding how to incorporate these parameters in expressions for $v_p^*$ and $v_\mathrm{rms}$ is likely to be non-trivial. For example, a scaling based on the mean free path of a fine particle, $c_3d-d_f = d(c_3-1/R)$, collapses the data in Figs.~\ref{fig:fig3}(a) and (c), but cannot be rigorously justified.  

Finally, our results apply to the single-fine-particle limit, but extending the key conclusions to binary mixtures with finite fine-particle concentrations, $c_f$, would also be valuable. Preliminary heap flow simulations with $R>4$ and global $c_f$ up to $30$\%, exhibit high-shear regions with local $c_f<5$\% where insights from the single-fine-particle limit are likely applicable. However, in low-shear regions, fine particles pack densely around bed particles, forming a continuous fine-particle phase that greatly reduces their vertical mobility.

\begin{acknowledgments}
We thank John P.\ Hecht, Alexander M.\ Fry, J\"{o}rg Theuerkauf, and Yi Fan for insightful discussions. This material is based upon work supported by the National Science Foundation under Grant No. CBET-2203703. 
\end{acknowledgments}

\bibliography{references}
\end{document}